\documentclass[twocolumn,draft,showkeys,showpacs,eqsecnum,nofootinbib,aps]{revtex4}
\def\beq{\begin{equation}}
\def\eeq{\end{equation}}
\def\bea{\begin{eqnarray}}
\def\eea{\end{eqnarray}}\def\nn{\nonumber}

\def\na{\nabla}
\def\pa{\partial}

\def\nn{\nonumber}

\begin{document}
\title{Singularities in geodesic surface congruence}
\author{Yong Seung Cho}
\email{yescho@ewha.ac.kr} \affiliation{Department of Mathematics,
Ewha Womans University, Seoul 120-750, Republic of
Korea}\affiliation{National Institute for Mathematical Sciences,
385-16 Doryong, Yuseong, Daejeon 305-340,  Republic of Korea}
\author{Soon-Tae Hong}
\email{soonhong@ewha.ac.kr} \affiliation{Department of Science
Education and Research Institute for Basic Sciences, Ewha Womans
University, Seoul 120-750 Republic of Korea}
\date{\today}
\begin{abstract}
In the stringy cosmology, we investigate singularities in geodesic
surface congruences for the time-like and null strings to yield
the Raychaudhuri type equations possessing correction terms
associated with the novel features owing to the strings. Assuming
the stringy strong energy condition, we have a Hawking-Penrose
type inequality equation. If the initial expansion is negative so
that the congruence is converging, we show that the expansion must
pass through the singularity within a proper time. We observe that
the stringy strong energy conditions of both the time-like and
null string congruences produce the same inequality equation.
\end{abstract}
 \maketitle


In 1970 Hawking and Penrose announced a celebrated result, the so
called Big Bang singularity theory~\cite{hawking70}. They
considered the congruence of geodesics in a spacetime manifold and
investigated the expansion, shear and twist of the time-like, and
null geodesic congruences.  The geodesic deviation equation yields
equations for their rate of change as one moves along the
geodesics.  The rate of change of the expansion gives us an
important equation known as the Raychaudhuri equation which
contains a curvature term.  They used the Einstein equation to the
curvature term in the equation.  If the tangent vector fields of
the geodesics in the congruence are hypersurface orthogonal, the
strong energy condition holds in the spacetime, and the expansion
takes the negative initial value at any point on a geodesic in the
congruence, then the expansion is shown to have a singularity on
the geodesic within a proper time.

Quite recently, we studied the variation of the surface spanned by
strings in a spacetime manifold~\cite{cho07}.  Using the
Nambu-Goto string action~\cite{nambu70,goto71}, we produced the
geodesic surface equation and the geodesic surface deviation
equation which yields a Jacobi field. Exploiting symplectic
cut-and-gluing formulae of the relative Gromov-Witten invariants,
one of us obtained a recursive formula for the Hurwitz number of
triple ramified geodesic surface coverings of a Riemann surface by
a Riemann surface~\cite{cho08}.

String theory emerged in 1968 when Veneziano proposed that the
relativistic scattering amplitudes of hadrons obey Regge
asymptotics and crossing symmetry associated with the Euler beta
function~\cite{vene68}. Scherk and Schwarz studied the small slope
expansion of the closed string theory to find that, at the zero
mass level, the theory contains a graviton~\cite{schwarz74}.
Gross, Harvey, Rohm, and Martin constructed heterotic string
theory by combining the closed 26-dimensional bosonic and
10-dimensional fermionic strings~\cite{gross85}. Candelas,
Strominger, Horowitz, and Witten~\cite{witten85} investigated
vacuum configurations in the superstring theory to produce
unbroken supersymmetry in four dimensions by compactifying the
extra six dimensional Calabi-Yau manifold~\cite{calabi57,yau77},
which is Ricci-flat and K\"ahler and possesses SU(3) holonomy.
Polchinski studied Dirichlet branes defined by mixed
Dirichlet-Neumann boundary conditions to yield electric and
magnetic Ramond-Ramond charges~\cite{pol95}, and Ho$\check{\rm
r}$ava and Witten proposed that the 10-dimensional heterotic
string theory is related to an 11-dimensional theory on an
orbifold~\cite{witten96}. Maldacena conjectured that the
compactifications of string theory on Anti-de Sitter spacetime is
dual to conformal field theories~\cite{malda97}.

Recently, the experimental data of the accelerating expansion of
the universe has suggested a small, positive vacuum expectation
value of the cosmological constant~\cite{perlmutter99}. Kachru,
Kallosh, Linde, and Trivedi constructed vacua of string theory to
yield a positive cosmological constant by introducing
supersymmetry breaking quantum corrections~\cite{kallosh03}.
Ooguri, Strominger, and Vafa conjectured that a supersymmetric
partition function for a four dimensional
Bogomolny-Prasad-Sommerfield black hole in a Calabi-Yau
compactification of superstring is related to a second quantized
topological string partition function associated with the black
hole charges~\cite{ooguri04}. The string theory has given us a
better understanding of the universe and the black holes in
cosmology and may provide an analytical tool for studying the
nature of the Big Bang singularity theory.


We consider a fibration $\pi:M\rightarrow N$ over a spacetime four
manifold $N$ with a $D$-dimensional total manifold $M$ associated
with the metric $g_{ab}$ and a Calabi-Yau manifold $F$ as a fiber
space. In analogy of the relativistic action of a point particle
in $N$, the action for a string is proportional to the area of the
surface spanned in the total manifold $M$ by the evolution along
the time direction of the string in $F$. We first introduce a
smooth congruence of time-like geodesic surfaces in $M$. We
parameterize the surface generated by the evolution of a time-like
string by two world sheet coordinates $\tau$ and $\sigma$, and
then we have the corresponding vector fields
$\xi^{a}=(\pa/\pa\tau)^{a}$ and $\zeta^{a}=(\pa/\pa\sigma)^{a}$.
Since we have gauge degrees of freedom, we can choose the
orthonormal gauge~\cite{scherk75,witten87,pol98} $\xi\cdot\zeta=0$
and $\xi\cdot\xi+\zeta\cdot\zeta=0$. In the orthonormal gauge, we
introduce tensor fields $B_{ab}$ and $\bar{B}_{ab}$ defined as
\beq B_{ab}=\na_{b}\xi_{a},~~~
\bar{B}_{ab}=\na_{b}\zeta_{a},\label{babbbabt}\eeq which satisfy
the identities $B_{ab}\xi^{a}=\bar{B}_{ab}\zeta^{a}=0$ and
$-B_{ab}\xi^{b}+\bar{B}_{ab}\zeta^{b}=0$. Here we have used the
geodesic surface equation
$-\xi^{a}\na_{a}\xi^{b}+\zeta^{a}\na_{a}\zeta^{b}=0$~\cite{scherk75,cho07}.
If the time-like curves of the geodesic surfaces are geodesic,
then the string curves are also geodesic.

We introduce the deviation vector field $\eta^{a}=(\pa/\pa
\alpha)^{a}$ which represents the displacement to an
infinitesimally nearby world sheet, and we consider the three
dimensional submanifold spanned by the world sheets. We then may
choose $\tau$, $\sigma$, and $\alpha$ as coordinates of the
submanifold to yield the commutator relations
$\pounds_{\xi}\eta^{a}=\pounds_{\zeta}\eta^{a}=\pounds_{\xi}\zeta^{a}=0$.
Using the above relations, we obtain
$\xi^{a}\na_{a}\eta^{b}-\zeta^{a}\na_{a}\eta^{b}=(B^{b}_{~a}-\bar{B}^{b}_{~a})\eta^{a}$.
Next we define the metrics $h_{ab}$ and $\bar{h}_{ab}$: \beq
h_{ab}=g_{ab}+\xi_{a}\xi_{b},~~~\bar{h}_{ab}=g_{ab}-\zeta_{a}\zeta_{b}.
\label{projections}\eeq Here one notes that $h_{ab}$ and
$\bar{h}_{ab}$ are the metrics on the hypersurfaces orthogonal to
$\xi^{a}$ and $\zeta^{a}$, respectively. We split $B_{ab}$ into
three pieces \beq B_{ab}=\frac{1}{D-1}\theta
h_{ab}+\sigma_{ab}+\omega_{ab},\label{bab}\eeq where the
expansion, shear and twist~\cite{hawking70,wald84,carroll} of the
stringy congruence along the time direction are defined as
$\theta=B^{ab}h_{ab}$, $\sigma_{ab}=B_{(ab)}-\frac{1}{D-1}\theta
h_{ab}$ and $\omega_{ab}=B_{[ab]}$. Similarly, as $B_{ab}$ in
(\ref{bab}) we can decompose $\bar{B}_{ab}$ into three parts; the
expansion, shear and twist of the stringy congruence along the
string coordinate $\sigma$-direction which are defined as
$\bar{\theta}=\bar{B}^{ab}\bar{h}_{ab}$,
$\bar{\sigma}_{ab}=\bar{B}_{(ab)}-\frac{1}{D-1}\bar{\theta}
\bar{h}_{ab}$ and $\bar{\omega}_{ab}=\bar{B}_{[ab]}$,
respectively.

Taking an ansatz that the expansion $\bar{\theta}$ is constant
along the $\sigma$-direction, one obtains a Raychaudhuri type
equation, namely, an evolution equation for the expansion \bea
\frac{d\theta}{d\tau}&=&-\frac{1}{D-1}(\theta^{2}-\bar{\theta}^{2})
-\sigma_{ab}\sigma^{ab}+\bar{\sigma}_{ab}\bar{\sigma}^{ab}\nn\\
&&+\omega_{ab}\omega^{ab} -\bar{\omega}_{ab}\bar{\omega}^{ab}
-R_{ab}(\xi^{a}\xi^{b}-\zeta^{a}\zeta^{b}).\label{treq} \eea We
now assume $\omega_{ab}=\bar{\omega}_{ab}$,
$\sigma_{ab}\sigma^{ab}\gg\bar{\sigma}_{ab}\bar{\sigma}^{ab}$ and
a stringy strong energy condition $R_{ab}(\xi^{a}\xi^{b}-\zeta^{a}\zeta^{b})\ge 0$
where \beq
R_{ab}(\xi^{a}\xi^{b}-\zeta^{a}\zeta^{b})=8\pi\left(T_{ab}
(\xi^{a}\xi^{b}-\zeta^{a}\zeta^{b})+\frac{2}{D-2}T\right),
\label{strong}\eeq and $T_{ab}$ and $T$ are the energy-momentum
tensor and its trace, respectively. The Raychaudhuri type equation (\ref{treq})
then has a solution of the form \beq \frac{1}{\theta}\ge
\frac{1}{\theta_{0}}+\frac{1}{D-1}\left(\tau-\int_{0}^{\tau} {\rm
d}\tau~\left(\frac{\bar{\theta}}{\theta}\right)^{2}\right),\label{soln}
\eeq where $\theta_{0}$ is the initial value of $\theta$ at
$\tau=0$. We assume that $\theta_{0}$ is negative so that the
congruence is initially converging as in the point particle case
shown below. The inequality (\ref{soln}) implies that $\theta$
must pass through the singularity within a proper time \beq
\tau\le \frac{D-1}{|\theta_{0}|}+\int_{0}^{\tau} {\rm
d}\tau~\left(\frac{\bar{\theta}}{\theta}\right)^{2}.\label{propertime}
\eeq For a perfect fluid, the energy-momentum tensor given by
$T_{ab}=\rho~u_{a}u_{a}+P~(g_{ab}+u_{a}u_{b})$ where $\rho$ and
$P$ are the mass-energy density and pressure of the fluid as
measured in its rest frame, respectively, and $u^{a}$ is the
time-like $D$-velocity in its rest frame~\cite{mtw,wald84}, the
stringy strong energy condition (\ref{strong}) yields only one inequality equation
\beq
\frac{D-4}{D-2}\rho+\frac{D}{D-2}P>0.
\label{secstring}
\eeq

Here one notes that, if the fiber space $F$ in our fibration $\pi:
M\rightarrow N$ is a point, then the total space $M$ is the same
as the base spacetime four manifold $N$.  In this case, the
geodesic surfaces are geodesic in $N$, the congruence of time-like
geodesic surfaces is a congruence of time-like geodesics, and so
$\bar{B}_{ab}=\bar{\theta}=\bar{\sigma}_{ab}=\bar{\omega}_{ab}=0$.
If the congruence is hypersurface orthogonal, then we have
$\omega_{ab}=0$.  Suppose that the strong energy condition
$R_{ab}\xi^{a}\xi^{b}\ge 0$ is satisfied to yield two
inequalities~\cite{hawking70,wald84,carroll} \beq
\rho+3P\ge0,~~~\rho+P\ge0.\label{strongpoint}\eeq We then have the
differential inequality equation
$\frac{d\theta}{d\tau}+\frac{1}{3}\theta^{2}\le 0$, which has a
solution in the following form: \beq \frac{1}{\theta}\ge
\frac{1}{\theta_{0}}+\frac{1}{3}\tau.\label{solnpoint} \eeq If we
assume that $\theta_{0}$ is negative, the expansion $\theta$ must
go to the negative infinity along that geodesic within a proper
time $\tau\le 3/|\theta_{0}|$.  This consequence coincides with
the one of Hawking and Penrose~\cite{hawking70}.


Next, we investigate the congruence of the null strings, where the
tangent vector of a null curve is normal to itself.  We consider
the evolution of vectors in a $(D-2)$-dimensional subspace of
spatial vectors normal to the null tangent vector field
$k^{a}=(\pa/\pa\lambda)^{a}$ where $\lambda$ is the affine
parameter, and to an auxiliary null vector $l^{a}$ which points in
the opposite spatial direction to $k^{a}$, normalized by
$l^{a}k_{a}=-1$~\cite{carroll} and is parallel transported,
namely, $k^{a}\na_{a}l^{b}=0$. The spatial vectors in the
$(D-2)$-dimensional subspace are then orthogonal to both $k^{a}$
and $l^{a}$.

We now introduce the metrics $n_{ab}$ defined below and
$\bar{h}_{ab}$ defined in (\ref{projections}), \beq
n_{ab}=g_{ab}+k_{a}l_{b}+l_{a}k_{b}.\label{projection2}\eeq
Similarly to the time-like case, we introduce tensor fields
$B_{ab}=\na_{b}k_{a}$ and $\bar{B}_{ab}$ in (\ref{babbbabt})
satisfying the identities $B_{ab}k^{a}=\bar{B}_{ab}\zeta^{a}=0$
and $-B_{ab}k^{b}+\bar{B}_{ab}\zeta^{b}=0$. We also define the
deviation vector $\eta^{a}=(\pa/\pa \alpha)^{a}$ representing the
displacement to an infinitesimally nearby world sheet so that we
can choose $\lambda$, $\sigma$, and $\alpha$ as coordinates of the
three dimensional submanifold spanned by the world sheets. We then
have the commutator relations
$\pounds_{k}\eta^{a}=\pounds_{\zeta}\eta^{a}=\pounds_{k}\zeta^{a}=0$
and
$k^{a}\na_{a}\eta^{b}-\zeta^{a}\na_{a}\eta^{b}=(B^{b}_{~a}-\bar{B}^{b}_{~a})\eta^{a}$.

We decompose $B_{ab}$ into three pieces \beq
B_{ab}=\frac{1}{D-2}\theta
n_{ab}+\sigma_{ab}+\omega_{ab},\label{babnull2}\eeq where the
expansion, shear and twist of the stringy congruence along the
affine direction are defined as $\theta=B^{ab}n_{ab}$,
$\sigma_{ab}=B_{(ab)}-\frac{1}{D-2}\theta n_{ab}$ and
$\omega_{ab}=B_{[ab]}$. It is noteworthy that even though we have
the same notations for $B_{ab}$, $\theta$, $\sigma_{ab}$ and
$\omega_{ab}$ in (\ref{bab}) and (\ref{babnull2}), the differences
of these notations among the time-like sting cases and null string
cases are understood in the context. Similarly we decompose
$\bar{B}_{ab}$ into three parts as in the time-like case. Taking
the ansatz that the expansion $\bar{\theta}$ is constant along the
$\sigma$-direction as in the time-like case, we have another
Raychaudhuri type equation \bea
\frac{d\theta}{d\lambda}&=&-\frac{1}{D-2}\theta^{2}+\frac{1}{D-1}\bar{\theta}^{2}
-\sigma_{ab}\sigma^{ab}+\bar{\sigma}_{ab}\bar{\sigma}^{ab}\nn\\
&&+\omega_{ab}\omega^{ab}-\bar{\omega}_{ab}\bar{\omega}^{ab}
-R_{ab}(k^{a}k^{b}-\zeta^{a}\zeta^{b}).\label{treqnull} \eea
Assuming $\omega_{ab}=\bar{\omega}_{ab}$,
$\sigma_{ab}\sigma^{ab}\gg\bar{\sigma}_{ab}\bar{\sigma}^{ab}$ and
a stringy strong energy condition for null case
$R_{ab}(k^{a}k^{b}-\zeta^{a}\zeta^{b})\ge 0$ which, exploiting the
energy-momentum tensor of the perfect fluid, reproduces the
inequality (\ref{secstring}) in the time-like congruence of
strings. The Raychaudhuri type equation (\ref{treqnull}) for the
null strings then has a solution in the following form: \beq
\frac{1}{\theta}\ge
\frac{1}{\theta_{0}}+\frac{1}{D-2}\left(\lambda-\frac{D-2}{D-1}\int_{0}^{\lambda}
{\rm
d}\lambda~\left(\frac{\bar{\theta}}{\theta}\right)^{2}\right),\label{solnnull}
\eeq where $\theta_{0}$ is the initial value of $\theta$ at
$\lambda=0$. We assume again that $\theta_{0}$ is negative. The
inequality (\ref{solnnull}) then implies that $\theta$ must pass
through the singularity within an affine length \beq \lambda\le
\frac{D-2}{|\theta_{0}|}+\frac{D-2}{D-1}\int_{0}^{\lambda} {\rm
d}\lambda~\left(\frac{\bar{\theta}}{\theta}\right)^{2}.
\label{propertimenull} \eeq

In the point particle limit with
$\bar{B}_{ab}=\bar{\theta}=\bar{\sigma}_{ab}=\bar{\omega}_{ab}=0$
and $\omega_{ab}=0$, we assume that the strong energy condition $R_{ab}k^{a}k^{b}\ge
0$ is satisfied to yield the second inequality of
(\ref{strongpoint})~\cite{hawking70,wald84,carroll}. If we assume
that the initial value is negative, the expansion $\theta$ must go
to the negative infinity along that geodesic within a finite
affine length~\cite{hawking70}.


We now have several comments to address. In (\ref{soln}),
(\ref{propertime}), (\ref{solnnull}) and (\ref{propertimenull}),
one notes that the correction terms associated with
$(\bar{\theta}/\theta)^{2}$ are the novel features of the stringy
congruence. Moreover, taking the ansatz
$\bar{\theta}^{2}\ll\theta^{2}$, which indicates that the internal
expansion $\bar{\theta}$ along the $\sigma$-direction is
negligibly small compared to the expansion $\theta$ along the time
(or affine) direction, the results (\ref{soln}) and
(\ref{solnnull}) reduce into the point particle results in the
$D=4$ limit, respectively. This observation does not contradict
the fact that the internal size of the string remains much less
than the Planck length $l_{P}=1.61\times 10^{-33}$ cm, as believed
in the string theory~\cite{witten87,pol98}.

Moreover, the $D=4$ limit of the stringy strong energy condition
(\ref{secstring}) for both the time-like and null cases does not
reduce to the well-known strong energy conditions in
(\ref{strongpoint}) for the $D=4$ point particle cases since the
stringy strong energy conditions have the additional contributions
$R_{ab}(-\zeta^{a}\zeta^{b})$ originating from the stringy degrees
of freedom. It is also remarkable to see that the stringy strong
energy conditions of both the time-like and null string
congruences produce for the perfect fluid the same result in
(\ref{secstring}), in contrast to the fact that the corresponding
strong energy conditions of the point particle congruences have
different forms, as shown in (\ref{strongpoint}). We, thus,
conclude that, in the higher $D$-dimensional stringy congruence
cosmology, both the massless gauge particles, such as photons and
gravitons, and the massive particles can be created in the same
cosmological environment described by the stringy strong energy
condition in (\ref{secstring}). In the point particle standard
cosmology in four dimensions, it is well known that the
radiation-to-matter transition exists and the radiation-dominated
phase precedes the matter-dominated one. However, if one follows
the above conclusion originated from the $D$-dimensional stringy
congruence theory, this stringy cosmology could not demand such
radiation-to-matter transition, and, thus, the radiation and the
matter can coexist in the same epoch along the evolution of the
universe after the Big Bang without any preference of the
dominated phases.

In the higher $D$-dimensional stringy congruence theory, one can
have the condition $\omega_{ab}=\bar{\omega}_{ab}$ associated with
the additional $\bar{\omega}_{ab}$. First, in the case of
$\omega_{ab}=\bar{\omega}_{ab}=0$, we can have the Hawking and
Penrose limit with $\omega_{ab}=0$ in the $D=4$ point particle
congruence cosmology~\cite{hawking70}. Second, we can additionally
have the $\omega_{ab}=\bar{\omega}_{ab}\neq0$.  In this case, one
can have the nonvanishing $\omega_{ab}$ initiate the desirable
rotational degrees of freedom encountered in the universe such as
the rotational motions of galaxies, stars, and planets.  Moreover
the nonvanishing $\bar{\omega}_{ab}$ could explain the rotational
degrees of freedom of the strings
themselves~\cite{scherk75,witten87,pol98}.

\acknowledgments The work of YSC and STH was supported by the
Korea Research Council of Fundamental Science and Technology
(KRCF), Grant No. C-RESEARCH-2007-11-NIMS, and the work of STH was
supported by the Korea Research Foundation (MOEHRD), Grant No.
KRF-2006-331-C00071.

\end{document}